\documentclass{article}
\usepackage{amssymb}

\usepackage{amsmath}
\usepackage{graphicx}

\def\Z{{\mathbb Z}}
\def\R{{\mathbb R}}

\def\d{{\mathrm d}}
\def\e{{\mathrm e}}
\def\c{{\mathrm{c}}}

\begin{document}
\begin{center}
{\huge Normal transport properties for a \\ classical particle coupled to \\ \vglue 0.1cm a non-Ohmic bath}
\\
\vspace*{20pt} P. {\sc Lafitte}$^{(1,2)}$, P.~E. {\sc
Parris}$^{(3)}$ and S. {\sc De Bi\`evre}$^{(1,2)}$
\\
\vspace*{5pt}
{\small
$^{(1)}$ Laboratoire Paul Painlev\'e, CNRS, UMR 8524 et UFR de Math\'ematiques
\\
Universit\'e des Sciences et Technologies de Lille
\\
F-59655 Villeneuve d'Ascq Cedex, France.
}

\vspace*{5pt}
{\small $^{(2)}$ Equipe-Projet SIMPAF
\\
Centre de Recherche INRIA Futurs
\\
Parc Scientifique de la Haute Borne,
40, avenue Halley B.P. 70478
\\
F-59658 Villeneuve d'Ascq cedex, France.
\\
E-mails:  {\tt  pauline.lafitte@math.univ-lille1.fr, Stephan.De-Bievre@math.univ-lille1.fr}
   }

\vspace*{5pt}
{\small $^{(3)}$ Department of Physics\\ Missouri University of Science and
Technology,\\ Rolla, MO 65409, USA\\
E-mail: {\tt parris@mst.edu}}
\\\vskip 0.1cm
{January 29 2008}
\end{center}

\begin{abstract}
We study the Hamiltonian motion of an ensemble of unconfined classical
particles driven by an external field $F$ through a
translationally-invariant, thermal array of \emph{monochromatic} Einstein
oscillators. The system does not sustain a stationary state, because the oscillators cannot effectively absorb the energy of high speed particles. We nonetheless show that the system has at all positive temperatures a well-defined low-field mobility $\mu $ over macroscopic time scales of order $\exp(-c/F)$. The mobility is independent of $F$ at low fields, and related to the zero-field diffusion constant $D$ through the Einstein relation. The system therefore exhibits normal transport even though the bath obviously has a discrete frequency spectrum (it is simply monochromatic) and is therefore highly non-Ohmic. Such features are usually associated with anomalous transport properties.
\end{abstract}


\section{Introduction}

The study of classical and quantum dynamical systems of a few degrees of freedom coupled to an environment containing many continues to attract considerable attention for various reasons. First of all, such studies shed light on fundamental issues of non-equilibrium statistical mechanics, such as return to equilibrium, the emergence of irreversibility and the existence of stationary non-equilibrium states. More generally, they arise in the rigorous derivation of macroscopic transport laws (such as Ohm's law and Fourier's law) and of reduced equations of motion (e.g., Langevin equations and master equations) from a microscopic, Hamiltonian description of the dynamics. The explicit computation of transport coefficients such as the diffusion constant or the mobility from first principles is another motivation of such studies, particularly in solid state physics. It is this problem that interests us here.
	
The case in which the environmental variables to which the particle couples are \emph{vibrational} degrees of freedom is of obvious interest to many fields of physics and is the most amenable to both analytical and numerical studies. The model Hamiltonian we will study in this paper was introduced in \cite{spd} and can be written
\begin{eqnarray}
H_{F} &=&\frac{1}{2}p^{2}-Fq+\sum_{m}\frac{1}{2}\left( p_{m}^{2}+\omega
^{2}q_{m}^{2}\right)  +\alpha\sum_{m} q_{m}\rho\left(
q-ma\right) .  \label{clashol}
\end{eqnarray}%
Here $q$ and $p$ are the particle position and momentum and $(q_m, p_m)$ are the position and momentum of an oscillator of frequency $\omega$ located at $x_m=ma$ with which the particle interacts provided $|q-ma|<\sigma$;  indeed, $\rho$ vanishes when $|q-ma|\geq \sigma$ and equals unity otherwise (we take $a-2\sigma \equiv L>0$). Here $F\geq 0$ is a driving field and we are interested in computing the low field mobility of the particle as a function of the temperature of the oscillator bath.

Our motivation for studying this particular model is twofold. On the one hand, the model obviously resembles those describing electron-phonon interactions in solids \cite{ho,ziman}, an analogy that is further worked out in \cite{dblp}. Alternatively it can be viewed as describing the motion of a particle in an inelastic one-dimensional
Lorentz gas.  Recall that in the usual Lorentz gas, a popular model for the study of transport properties, a particle moves on a two-dimensional plane on which  circular hard scatterers are randomly or periodically placed. In order to describe the interaction with a thermal environment, the scatterers can be allowed to rotate, as in \cite{ey, llm}, yielding a model similar to ours but with the environment described through rotational degrees of freedom rather than vibrational ones. Alternatively, the scatterers can be kept fixed, as in \cite{cels1, cels2}, while the environmental interaction is phenomenologically described through the use of a Gaussian thermostat; in the latter situation only the particle degrees of freedom remain dynamical and the system is no longer Hamiltonian.
Our model can be viewed as a one-dimensional Hamiltonian version of such models. Unlike previously studied two-dimensional transport models, however, the present model is not homogeneous in energy; as a result it displays a wide range of microscopic behavior that varies with the temperature of the system.

The main conclusion of the paper regarding the present model is that, provided the driving field $F$ is low enough, the particle
distribution quickly acquires a drift velocity $%
v_{F}$ proportional to the driving field $F$ which it then maintains over times that are exponentially long in $F^{-1}.$
This current is  \emph{metastable}, however, due to a runaway phenomenon that arises because oscillator lattices, in general, do not efficiently slow down very fast particles (see Section~\ref{s:runaway}). Thus, this system does not approach a stationary non-equilibrium state. We nevertheless show that
the associated mobility is well-defined and related by the
fluctuation-dissipation theorem to the diffusion constant $D$ and the
inverse temperature $\beta =1/kT$ through Einstein's relation $\mu =\beta D.$ Thus, in this simple fully Hamiltonian model, the particle exhibits normal transport behavior obeying Ohm's law at low fields.

These conclusions will be surprising to some, since previous studies of similar models seem to indicate rather strongly that coupling to a \emph{monochromatic} bath would be expected to lead to anomalous particle transport. To support this claim we conclude this introduction with a short overview of the relevant literature.
There is a considerable body of work on a variety of models in which a particle of momentum $p$
and position $q$ couples linearly to the displacements or momenta of a heat bath of oscillators. A very general class of such models was proposed in \cite{zw}. To describe the results in the literature relevant to our work here it will  be sufficient to consider a specific subclass of models described by the following Hamiltonian \cite{bdb}:
\begin{eqnarray}\label{generalH}
H(q,p;X)&=&\frac12 p^2 +V(q) \nonumber \\
&\ &\qquad +\frac12\int_D\int_B \left( p(x,\xi)^2+\omega(\xi)^2 q(x,\xi)^2\right) \sigma_1(x)\sigma_2(\xi) \d x \d \xi  \nonumber\\ &\ &\qquad \quad +\int_D\int_B \rho(x-q) c(\omega(\xi)) q(x,\xi)\sigma_1(x)\sigma_2(\xi) \d x \d \xi.
\end{eqnarray}
Here $X=(q(x,\xi), p(x, \xi))_{x\in D\subset\R, \xi\in B\subset \R^d}$ where $q(x, \xi)$, $p(x, \xi)$ are the displacement and momentum of an oscillator of frequency $\omega(\xi)$, which should be thought of as finding itself at the point $x\in D$. The parameter $\xi$ is to be thought of solely as an index for the oscillators. As a result, the collection of oscillators $(q(x, \xi), p(x,\xi))_{\xi\in B\subset \R^d}$, $x$ being fixed, represent a bath of vibrational degrees of freedom of the medium at the point $x\in D$.  Furthermore $\sigma_1(x) \d x$ and $\sigma_2(\xi) \d \xi$ are measures (on $D$ and $B$ respectively) and $\rho(x), c(\omega(\xi))$ positive functions determining the coupling of the oscillators to the particle. Note that, since neither $\sigma_2$ nor $c$ depend on $x$, the oscillator baths have the same characteristics at all points $x\in D$. We will choose for $\rho$ a compactly supported positive function centered at the origin. The Hamiltonian (\ref{clashol}) is of the above type, with
$$
\sigma_1(x)=\sum_m \delta(x-ma), \quad \sigma_2(\xi)=\delta(\xi-\omega),
$$
and $\omega(\xi)=\omega$.

Previously studied models with Hamiltonians as in (\ref{generalH}) are of two types. Models with a continuous distribution of oscillators uniformly throughout all of space are studied in \cite{aw, bdb}; in that case $\sigma_1(x)=1$ and $D=\R$. Alternatively, in \cite{co1, co2}, the oscillators are located at a countable set of points $D=\{x_m | m\in\Z\}$ chosen randomly and homogeneously throughout space so that
\begin{equation*}
\sigma_1(x)=\sum_{m\in \Z} \delta(x-x_m).
\end{equation*}

To discuss the results obtained in those previous studies we introduce the coupling weighted spectral density of the system, defined as
\begin{equation*}
J(\omega)=\frac{\pi}{2}\frac{c(\omega)^2}{\omega}n(\omega),
\end{equation*}
where the spectral density $n(\omega)$ is defined via
\begin{equation*}
N(\omega)=\int_{\omega(\xi)\leq \omega} \sigma_2(\xi) \d \xi=\int_0^\omega n(\omega')\d \omega'.
\end{equation*}
In \cite{aw, bdb, co1, co2}, the distribution of frequencies is supposed to be continuous and  $J(\omega)\sim\omega^s$ as $\omega\to0$, for some $s>0$.  It is then argued in \cite{aw, co1, co2} that normal transport properties are obtained in such models only if
$s=1$. In particular, these authors argue that when $V(q)=0$, particle motion is diffusive provided $s=1$, whereas anomalous diffusion occurs otherwise. Similar results are obtained rigorously in \cite{bdb} where it is proven that in the fully translationally invariant models considered (namely $\sigma_1(x)=1$) a particle moving at constant speed $v$ through an oscillator bath at zero temperature experiences a friction force proportional to $v^s$ at low $v$: this again indicates that a finite mobility will be obtained only if $s=1$, a fact rigorously proven in \cite{bdb} for the model considered, at zero temperature.

Similar conclusions have been obtained in the much more numerous studies (\cite{cl1, cl2, cefm, ddll, fkm, jp4, lcdfgz} and references therein) dealing with a simpler but related class of models in which
the potential $V\left( q\right)$ is chosen to be confining. Note that this prevents the particle from making arbitrarily large excursions from
equilibrium, justifying a ``dipole approximation'' in the Hamiltonian (\ref{generalH}) in which the coupling is
linear in $q$ as well. After some rewriting, the resulting bilinearly coupled Hamiltonian takes the form%
\begin{eqnarray*}
H_{\text{CL}}(q, p, X) &=&\frac{p^{2}}{2}+V(q)+\frac{1}{2}\int_B(p(\xi)^2 +\omega(\xi)^2q(\xi)^2) \sigma_2(\xi) \d \xi\notag \\
&&\,\;\;\;\;\;\;\;\;\;\;\;\;\;\;\;\;\;+q\int_B c(\omega(\xi))q(\xi)\sigma_2(\xi)\d \xi%
\label{HCL}
\end{eqnarray*}%
and is then much simpler to analyze. In classical systems, for example, it
straightforwardly leads to a generalized Langevin equation for the particle:
\begin{equation*}\label{glang}
\ddot q(t)+\int_0^t \gamma(t-s)\dot q(s) \d s=-\nabla V_{\mathrm{eff}}(q(t)) +f(t),
\end{equation*}
with
\begin{equation*}
V_{\mathrm{eff}}(q)=V(q)-\frac12\left[\int_B\frac{c(\omega(\xi))^2}{\omega^2(\xi)}\sigma_2(\xi)\d\xi\right] q^2
\end{equation*}
and where
\begin{equation*}
\gamma(\tau)=\int_B\frac{\c(\omega(\xi))^2}{\omega^2(\xi)} \cos (\omega(\xi)\tau)\ \sigma_2(\xi)\d \xi=\frac{2}{\pi}\int_0^{+\infty} \frac{J(\omega)}{\omega} \cos(\omega \tau)\ \d \omega
\end{equation*}
is the retarded friction kernel and
\begin{equation}
 f(t)=-\int_B\c(\omega(\xi))\left[\left(q_0(\xi) +\frac{c(\omega)}{\omega^2} q_0)\right)\cos(\omega(\xi)t) + \pi_0(\xi)\sin(\omega(\xi)t)\right]
\sigma_2(\xi)\ \d\xi
\end{equation}
the random force exerted by the medium on the particle, which depends on the initial conditions for the particle ($q(0)=q_0$) and for the bath variables ($q_0(\xi), p_0(\xi)$).
Return to equilibrium for such systems was shown for general confining potentials in \cite{jp4} provided $s\leq 1$; the particular case $s=1$, with a harmonic potential $V$ was treated in \cite{cefm, gsi}. Note furthermore that the same model, with $V(q)=0$ can also be understood as a model for a non-confined particle: in that case the oscillator degrees of freedom can be thought of as internal degrees of freedom of a composite particle, rather than as environmental degrees of freedom.  For this model, it was shown in \cite{gsi, HA, sg} that the particle diffuses only in the Ohmic case $s=1$ and is superdiffusive (respectively subdiffusive) when $s>1$ (respectively $s<1$).

In short, in all existing literature in which theoretical results are available on the behavior of a particle coupled (linearly or not) to vibrational degrees of freedom, normal transport properties for the particle occur only if the oscillator bath has a continuous oscillator spectrum with a sufficient density of low frequency modes ($s=1$). Our study here shows that while an Ohmic local bath is certainly a sufficient condition for normal transport, it is not a necessary one. On the contrary, we show that even with a monochromatic oscillator bath having no low frequency modes, a finite diffusion constant and well behaved low field mobility can be obtained.

The paper is organized as follows. In Section~\ref{s:mobnum} we describe our numerical results on the particle mobility and its temperature dependence. Section~\ref{s:hightempmob} provides an explanation of the high temperature part of our data in terms of a random walk model motivated by the underlying physics of the Hamiltonian dynamics. In Section~\ref{s:linearresponse} we apply linear response theory to prove an Einstein relation for the model, valid at all temperatures, which we apply to explain the low temperature behaviour of the model. In Section~\ref{s:runaway} we show that no stationary state is established in the system due to a runaway phenomenon inherent in all models of this type. 


\begin{figure}
\centerline{
\includegraphics[height=6cm, keepaspectratio]{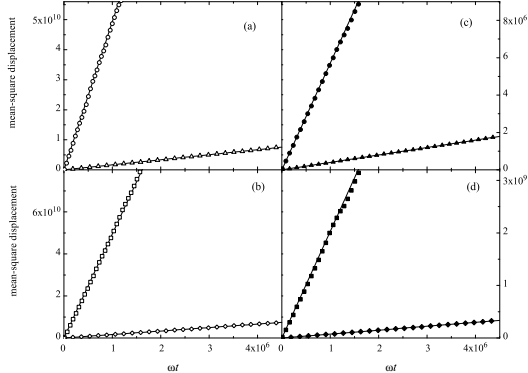}
}
\caption{Mean square displacement of the undriven particle ($F=0$) as a function
of $\protect\omega t$ for several sets of parameters and temperatures. In both panel (a) (where $c_{1}=0.015$, $c_{3}=0.5$) and panel (c) (where $%
c_{1}=0.5$, $c_{3}=0.5$) triangles indicate $c_{2}=0.5$ and circles $c_{2}=5.$
In both panel (b) (where $c_{1}=0.002$, $c_{3}=2,$) and panel (d) (where $c_{1}=0.07$, $c_{3}=2,$), 
diamonds indicate $c_{2}=0.5$ and squares $c_{2}=5$. Each
curve is an average of $10^{3}-10^{5}$ trajectories. Straight lines are
linear fits and $c_1, c_2, c_3$ are defined in (\protect\ref{eq:c1c2})-(\protect\ref{eq:c3}).\label{fig:diffusion}}
\end{figure}

\section{Mobility: numerical results}\label{s:mobnum}
A simple inspection of the equations of motion shows that, under the dynamics generated by the Hamiltonian \eqref{clashol}, the
particle undergoes at all times a constant acceleration $F$. Its motion is decoupled from the motion of the oscillators, except at those instants of time when it reaches the edge of one of the interaction regions, in other words when
$|q(t)-ma|=\sigma $ for some $m\in\Z$.  At those times, it encounters a potential energy step $\Delta $ of magnitude $\left| \alpha q_{m}\right| $ and it therefore either turns back or proceeds with a change in momentum that is simply determined by energy conservation. If the kinetic energy of the particle at the moment it arrives at the edge of the interaction region is greater than the potential discontinuity, it passes into the next region with a new kinetic energy reduced or increased by the potential step that it traverses in doing so. Otherwise it undergoes an elastic reflection. The oscillator at $x_m=ma$, on the other hand, oscillates about the displaced equilibrium $q_{m\ast }=-\alpha /\omega ^{2}$ at those times $t$ for which $|q(t)-ma|<\sigma $, and about the unperturbed equilibrium $q_{m}=0$ otherwise.

These features allow for an event-driven numerical implementation of the dynamics, without the need for numerical integration of a non-linear differential equation. As a result, we have been able to run simulations over very long times (of the order of $10^6-10^7$ oscillator periods), computing average displacements or mean square displacements for the particle, with an ensemble of initial conditions for the particle-oscillator system drawn from the (field-free) Boltzmann distribution. To obtain satisfactory statistics, we ran between $10^3$ and $10^5$ trajectories for each set of parameters considered.

In this paper we are concerned both with the high and the low temperature regime. To understand what this means, note that there are two mechanical energy scales associated with this model. The first one is the binding energy $-E_B=-\alpha ^{2}/2\omega ^{2}$, which is the classical ground state energy of the system, obtained when the particle is at rest in one interaction region (for example the one associated with the oscillator at $x=0$), with the corresponding oscillator in its displaced equilibrium position and all other oscillators at rest ($q_m=0=p_m, m\not=0$).  The other one is the kinetic energy $E_0=\sigma^2\omega^2$ of a particle that crosses the interaction region of length $2\sigma$ in one oscillator period. Along with the thermal energy $kT = 1/\beta$, this allows  the introduction of two energy related dimensionless parameters, which we denote by
\begin{equation}\label{eq:c1c2}
c_1\equiv\beta E_{B}\quad \mathrm{and}\quad 
c_{2}\equiv E_{B}/E_{0}.
\end{equation}
The model also has two length scales $L=a-2\sigma$ and $\sigma$, leading to a third dimensionless parameter
\begin{equation}\label{eq:c3}
c_{3}\equiv 2\sigma /L.
\end{equation}
By high temperatures we will mean the regime $c_1<< 1, \beta E_0=c_1/c_2<< 1$, for which the average thermal energy of the particle is (much) higher than both the binding energy $E_B$ and $E_0$. In the low temperature regime, these inequalities are reversed. As we will show below, the temperature dependence of the mobility is in this model qualitatively different in these two different regimes, due to the qualitatively different nature of the microscopic dynamics.

\begin{figure}[t]
\centerline{
\includegraphics[height=6.5cm, keepaspectratio]{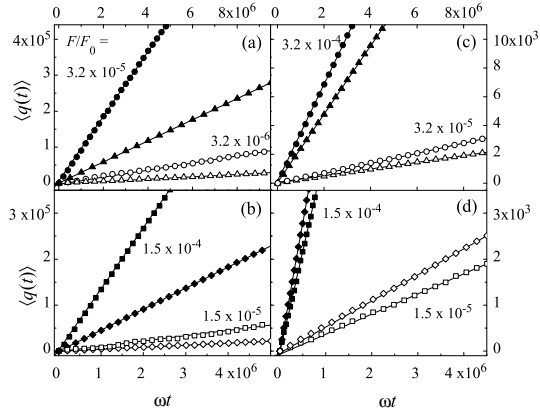}
}
\caption{Mean particle displacement as a function
of $\protect\omega t$ for several sets of parameters, temperatures, and
forces ($F_0=a\omega^2$). In both panel (a) (where $c_{1}=0.015$, $c_{3}=0.5$) and panel (c) (where $%
c_{1}=0.5$, $c_{3}=0.5$) triangles indicate $c_{2}=0.5$ and circles $c_{2}=5.$
In both panel (b) (where $c_{1}=0.002$, $c_{3}=2,$) and panel (d) (where $c_{1}=0.07$, $c_{3}=2,$),
diamonds indicate $c_{2}=0.5$ and squares $c_{2}=5$. Each
curve is an average of $10^{3}-10^{5}$ trajectories. Straight lines are
linear fits.\label{fig:drifts}}
\end{figure}

In Fig. \ref{fig:diffusion} plots are seen of the numerically computed mean square displacements of the particle for various parameter values, in absence of a driving field. As already observed in \cite{spd}, the motion is diffusive.

In Fig.~\ref{fig:drifts} we present plots of numerically calculated values of the averaged particle displacement $\langle q\left( t\right) \rangle $ in units of $a,$ in the presence of a driving field for several systems and temperatures.
In each case, the average is computed over initial conditions of the particle-chain system which are drawn from its joint thermal equilibrium distribution in absence of the driving field.
One observes that  $\langle q\left( t\right) \rangle $
is linear in $t$ over macroscopically long times and distances. Over the time scales explored, each system
investigated thus attains a well-defined drift speed $v_{F}$, that is easily
determined through linear regression. 

\begin{figure}[t]
\center \includegraphics[height=4.100in]{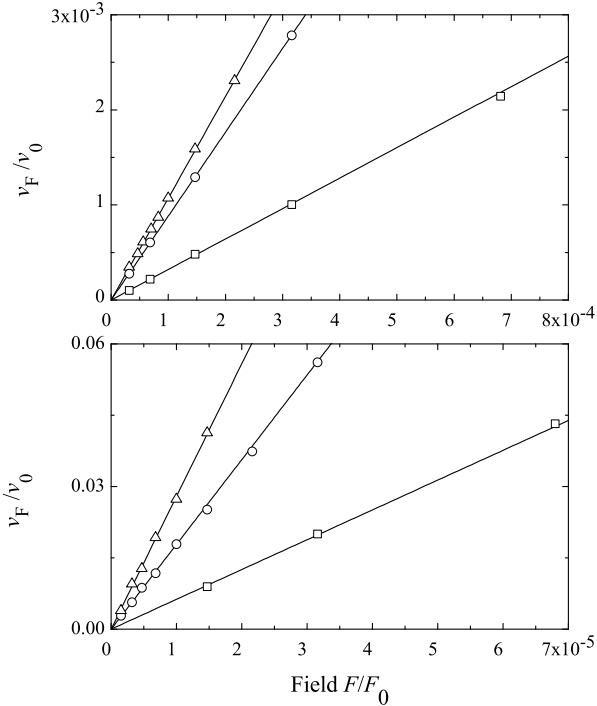}
\caption{Drift velocity $v_{F}$ as a
function of $F$. Straight lines are linear fits. For all, $c_{2}=5,$ $v_{0}=a%
\protect\omega $ and $F_{0}=a\protect\omega ^{2}.$ In the top (bottom) panel,
triangles indicate $c_{1}=0.5$ ($0.015$), $c_{3}=0.5$; squares indicate $%
c_{1}=0.9 $ ($0.03$), $c_{3}=1$; circles indicate $c_{1}=0.7$ ($0.02$), $c_{3}=2.$ \label{FIG2}}
\end{figure}

\begin{figure}[t]
\center \includegraphics[height=3in, keepaspectratio]{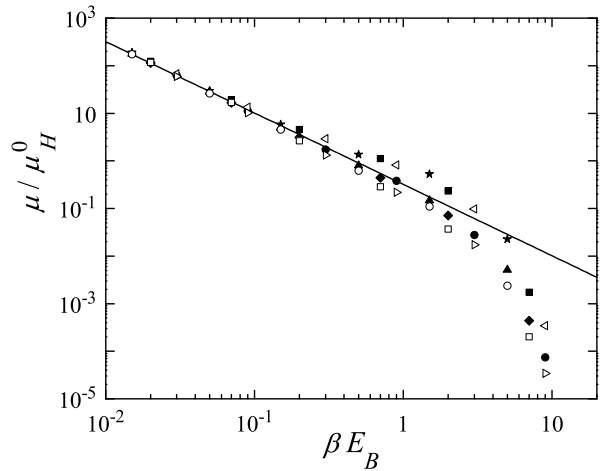}
\caption{Numerically computed mobilities $\protect\mu/\mu_H^0$ (see (\protect\ref{eq:muHighT}) for $\mu_H^0$) as a
function of $\protect\beta E_{B}$ over a wide range of temperatures and for nine different sets of model parameters, distinguished with different symbols. For each parameter set, the mobility is shown for six different temperatures. The straight line corresponds to $\protect\eta \left( \protect\beta %
E_{B}\right) ^{-3/2}$ with $\protect\eta =0.32$.\label{muHighT}}
\end{figure}

As seen in Fig.~\ref{FIG2}, for fixed $%
T$ and system parameters the resulting drift speeds are, for sufficiently
small fields, linear in $F$.
This allows us to extract through linear
regression, a well-defined low field mobility which we have plotted in  Fig.~\ref{muHighT} as a function of $\beta E_{B}$.

\begin{figure}[t]
\center
\includegraphics[height=6cm]{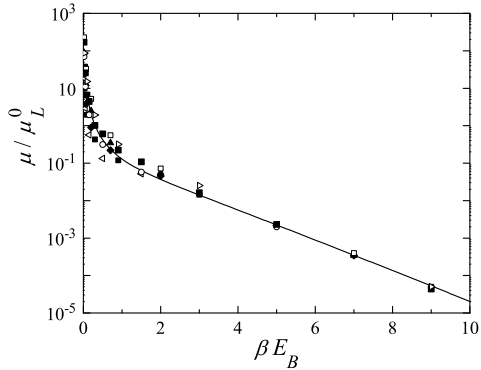}
\caption{A log-linear plot of  $\protect\mu/\mu_L^0 $ (see (\protect\ref{eq:muLowT}) for the definition of $\mu_L^0$) as a
function of $\protect\beta E_{B}$ for the same model parameters as in Fig.~\protect\ref{muHighT}, emphasizing the low temperature data. The curve corresponds to $0.14( x^{1/2} + 1.5x^{-3/2} )exp(-x)$ with
$x = \beta E_{B}$.
This clearly brings out both the
exponential dependence at low temperatures and the power law dependence at high temperatures. \label{fig:muLowT}}
\end{figure}

At low temperatures, the numerics is complicated by the following phenomenon.  First, since we compute displacements and since the system is periodic, we can start all particles in the cell at $q=0$. Then
$$
P_{\mathrm{int}}=\frac{2\sigma}{L\e^{-\beta E_B}+2\sigma},\quad P_{\mathrm{non}}=
\frac{L\e^{-\beta E_B}}{L\e^{-\beta E_B}+2\sigma}
$$
are the probabilities that the particle finds itself respectively in the interaction or in the non-interaction region of the cell \cite{spd}. Note that, obviously, the latter is exponentially less likely as the temperature is lowered. In addition, only an exponentially small fraction of the particles in the interaction region will ever leave the first cell under the dynamics, because most particles will never manage to overcome the typical energy barriers they meet at the edges. This leads to an exponential suppression of the mobility at low temperatures that can be observed both on Fig. \ref{muHighT} and Fig. \ref{fig:muLowT} and that will be further explained in Section
\ref{s:linearresponse}. These phenomena also affect the numerics in the following way: trapped particles will of course not contribute to the mobility and running the corresponding trajectories would be a waste of computer time. It would also keep one from getting good enough statistics, especially at low temperatures, where the drift speed is low. Fortunately, using the analysis of \cite{dbps}, it is possible to predict {\em a priori} from the initial conditions of the particle and the oscillator if the particle will ever get out, and implementing this in the program allows one to avoid running the corresponding trajectories uselessly. 
It follows from this analysis that particles can remain trapped only if they start in an interaction region. 
Suppose therefore $(q,p)$ are the initial data for the particle, with $0\leq q\leq 2\sigma$ so that the particle is indeed in the interaction region, and let $(q_0, p_0)$ be the initial conditions for the oscillator in the cell at the origin.  The total energy of this  particle-oscillator system can then be written
$$
E=\varepsilon(q) +\varepsilon_{\mathrm{osc}}(q_0, p_0)-E_B-2F\sigma,
$$
where
$$
\varepsilon_{\mathrm{osc}}(q_0, p_0)=\frac12(p_0^2 + (q_0+\alpha)^2)),\qquad \varepsilon(q)=\frac12 p^2-F(q-2\sigma).
$$
Analyzing the particle-oscillator dynamics along the lines of \cite{dbps} the necessary and sufficient conditions guaranteeing that the particle will leave the cell at the origin at some point in the future are
$$
2E_B-\varepsilon(q) \geq \varepsilon_{\mathrm{osc}}(q_0, p_0) > \frac{1}{4E_B}(\varepsilon(q)-2E_B)^2.
$$

We now turn in the next sections to the theoretical explanation of the observed numerical data described here and in particular of the temperature dependence of the computed mobilities observed in Fig.~\ref{muHighT} and Fig.~\ref{fig:muLowT}.


\section{High temperature mobility: a biased random walk model} \label{s:hightempmob}
In order to explain theoretically the high temperature part of the numerical results described in the previous section, we will further develop the simplified random walk model of the fully Hamiltonian dynamics at high temperature that was developed in \cite{spd} for the case $F=0$. This will provide us with a clear and simple physical picture of the motion of the particle in this regime. First note that, at high $T$, the typical kinetic energy of a particle (which is of order $kT$%
) is much higher than the typical barriers it encounters (which are of order $\sqrt{E_{B}kT}$). As a result, particles drawn from a thermal distribution tend to enter most interaction regions they
encounter, rather than reflecting. Now any particle with a high velocity $v$ will cross the interaction region in a time $\sigma/v$ which is short compared to the oscillator period and as a result it will tend
to lose an energy $\Delta E=-4E_{B}E_{0}/v^{2}$ to the oscillator, as shown in \cite{spd}.
At high temperatures, one has $\beta E_0<<1$ and a particle with approximate thermal speed will satisfy the above condition since
$
v_{\mathrm{th}}=\beta^{-1/2}
$
so that $\sigma/v_{\mathrm{th}}<<1/\omega$ is equivalent to $E_0\beta<<1$.

On the other hand, the same particle gains or loses an energy $Fa$ from the field while traversing the cell, depending on whether it is moving to the right or to the left. Consequently, if the field is too large, $Fa\geq 4E_{B}E_{0}/v_{\mathrm{th}}^{2}$, particles with thermal speed moving to the right will be accelerated indefinitely, so that it is not possible for a drift speed to be set up in the system. At small fields, on the other hand
\begin{equation}\label{eq:criticalfield}
Fa << 4E_{B}E_{0}/v_{\mathrm{th}}^{2}=4\beta E_B E_0
\end{equation}
such particles will lose  a net energy. The resulting average velocity change upon traversal of one cell is given by
\begin{equation}
\Delta v=\Delta _{F}E\frac{1}{v}=-\frac{4E_{B}E_{0}}{v^{3}}+\frac{1}{|v|}Fa.
\label{deltav}
\end{equation}%
So, when the temperature is high and the field small, the energy
loss experienced by most particles in the thermal distribution is small, and they traverse many cells before slowing down to a point where their kinetic energy is smaller than or comparable to the typical barrier heights. At that point, they can be expected to receive a randomizing ``kick'' from one of the oscillators. We will assume that this randomizing ``kick'' reinstates a thermal distribution for the particles.
One easily finds the
distance $\ell _{F}(v)$ over which a particle of high initial velocity $v$ travels before  slowing
down, as well as the characteristic time $\tau _{F}(v)$ of the slowing down process,  by integrating (\ref{deltav}):
$$
\ell_F(v)= \int_v^0 \frac{\d \ell}{\d v} \ \d v=\int_0^v \frac{v^3}{4E_BE_0}\frac{1}{1-\frac{Fa}{4E_BE_0}v|v|}\ \d v,
$$
and
$$
\tau_F(v)=  \int_v^0 \frac{\d t}{\d \ell}\frac{\d \ell}{\d v} \ \d v=\int_0^v\frac{1}{v} \frac{v^3}{4E_BE_0}\frac{1}{1-\frac{Fa}{4E_BE_0}v|v|}\ \d v.
$$
This yields:
\begin{equation*}
\tau _{F}(v)=\tau (v)\left[ 1+\frac{3}{5}\frac{Fav^{2}}{4E_{B}E_{0}}\frac{v}{%
|v|}\right] ,\qquad \tau \left( v\right) =\frac{\left| v^{3}\right| a}{%
12E_{B}E_{0}}
\end{equation*}%
and
\begin{equation*}
\ell _{F}(v)=\ell \left( v\right) \left[ 1+\frac{Fav^{2}}{6E_{B}E_{0}}\frac{v%
}{|v|}\right] ,\qquad \ell \left( v\right) =\frac{3}{4}v\tau \left( v\right).
\end{equation*}%
Note that, as expected, for $v>0$, one has $\ell_F(v)\geq \ell_F(-v)$: particles moving to the right are pushed on by the field and therefore travel further before slowing down.

The result of this analysis is that at high temperatures one can view the particle motion as a biased
random walk with
steps $\ell _{F}(v)$ and waiting times $\tau _{F}(v)$, with $v$ drawn after
each randomizing kick from the Boltzmann distribution $(\beta /2\pi
)^{1/2}\exp(-\beta v^{2}/2)$. This picture therefore leads one to predict an average drift velocity%
\begin{equation}\label{eq:vF}
v_{F}=\frac{\langle \ell _{F}\rangle }{\langle \tau _{F}\rangle }=\frac{Fa}{%
8E_{B}E_{0}}\frac{\langle v^{6}\rangle }{\langle \left| v\right| ^{3}\rangle
}=\mu F,
\end{equation}
and thus a mobility%
\begin{equation}
\mu =\frac{15a}{16}\sqrt{\frac{\pi }{2}}\sqrt{\frac{E_{B}}{E_{0}^{2}}}%
\;\left( \beta E_{B}\right) ^{-3/2}\equiv \mu _{H}^{0}\;\left( \beta
E_{B}\right) ^{-3/2}.\label{eq:muHighT}
\end{equation}%
The data presented in
Fig.~\ref{muHighT} show that this simple picture of the dynamics yields the correct temperature behaviour of the mobility at high temperatures as well as a reasonable estimate of the prefactor.

\section{Low temperature mobility: linear response}\label{s:linearresponse}

It was shown in \cite{spd} that the physics underlying the low-temperature dynamics of the model at zero field ($F=0$) is of a very different nature than the one at high temperatures. It can again be described in terms of a random walk but this time with \emph{nearest-neighbour} hopping of the particle.  The analysis of \cite{spd} yields a very good prediction for the diffusion constant and its temperature dependence at low $T$:
\begin{equation}
D\sim D_{\text{th}}^{0}\frac{\exp \left( -\beta E_{B}\right) }{\sqrt{\beta
E_{B}}}\qquad \qquad \beta E_{B}\gg 1  \label{eq:Dthermlow}
\end{equation}%
where%
\begin{equation}
D_{\text{th}}^{0}=\frac{a^{2}}{2\sigma }\sqrt{\frac{E_{B}}{2\pi }}.
\label{eq:Dlow0}
\end{equation}%
In particular, it is seen from this result that the diffusion constant is thermally activated due to the self-trapping of the particle at low temperatures briefly explained in Section \ref{s:mobnum} (see \cite{spd} for details). This low temperature analysis is however more involved than the high-temperature one and will not be further adapted to the non-zero field situation we consider here. Instead we will present in this section an alternative derivation of the mobility of the system and of its temperature dependence via Kubo's linear response theory. We will show this leads to a \emph{finite time} variant of the Einstein relation  valid at all temperatures and for small non-zero fields which, combined with (\ref{eq:Dthermlow}), yields for the low temperature behaviour of $\mu$ the result
\begin{equation}\label{eq:muLowT}
\mu\sim \mu^0_L\sqrt{\beta E_{B}} \exp \left( -\beta E_{B}\right),\qquad \mu_L^0=\frac{a^{2}}{2\sigma }\sqrt{\frac{1}{2\pi E_{B}}}.
\end{equation}
This is in very good agreement with the low temperature numerical data as can be seen in Fig.~\ref{fig:muLowT}.

Let us consider again the Hamiltonian in (\ref{clashol}) and write it as
$$
H_F(q,p; X)=H_0(q,p ; X) -Fq,
$$
where $X=(q_m,p_m)_{m\in\Z}$. We will write $\Phi_t^F$ for the phase space flow generated by this Hamiltonian and use the shorthand $\Phi_t^F(q,p;X)=(q(t; F), p(t; F))$. In other words, $q(t,F)$ and $p(t;F)$ are the position and  the momentum at time $t$ of a particle that at time $0$ had position $q$ and momentum $p$ and that evolved under the dynamics generated by $H_F$ in a medium with initial condition $X$. We will be interested in the thermally averaged mean velocity of the particle at time $t$,
$$
v_F(t)=\frac{\langle (q(t;F)-q)\rangle}{t},
$$
which is the thermal average of its mean displacement, divided by the time $t$. Also, we define the mobility by
$$
\mu_F(t) \equiv \frac{v_F(t)}{F}.
$$
Note that this is, a priori, a time and field dependent quantity and that $v_F(t)$ and $\mu_F(t)$ are precisely the quantities that we compute numerically.  We are interested in their long time and small field behaviour. Here and in what follows, for any phase space function $f(q,p ; X)$, $\langle f\rangle$  designates the thermal average of $f$ for the Hamiltonian $H_0$. In addition
$$
\Delta f_t = \langle f\circ \Phi_t^{F}\rangle -\langle f\rangle.
$$
A formal application of first order perturbation theory to our system then shows readily that for any phase space function $f$,
$$
\Delta f_t =\beta \int_0^t \d s\ F \langle\{q, H_0\} f\circ \Phi_{t-s}^0\rangle +{\mathrm O}_t(F^2).
$$
Consequently, if $f=q$, one finds, after a change of variables in the integral
$$
v_F(t) = \beta F\frac{1}{t}\int_0^t \d s\ \langle p q(s;0)\rangle+{\mathrm O}_t(F^2).
$$
 Writing furthermore
$$
q(s;0)-q=\int_0^s  \d s' p(s';0)
$$
this can be rewritten, after changing the order of the time integrations:
$$
v_F(t) = \beta F\int_0^t \d s'\ \left(1-\frac{s'}{t}\right)\langle p p(s';0)\rangle+{\mathrm O}_t(F^2).
$$
On the other hand, let us introduce the mean square displacement of a freely evolved particle in thermal equilibrium
$$
D(t):=\frac{\langle (q(t;0)-q)^2\rangle}{2t}.
$$
This can be rewritten in the usual way as
$$
D(t)=\int_0^t  (1-\frac{s}{t})\langle pp(s)\rangle\ \d s.
$$
Hence, we immediately obtain the following \emph{finite time} version of the Einstein relation:
\begin{equation}\label{eq:mobfinitet}
\mu_F(t)=\beta D(t) +{\mathrm O}_t(F),
\end{equation}
which holds for all times and fields.

\begin{figure}[t]
\begin{center}
 \includegraphics[height=6cm]{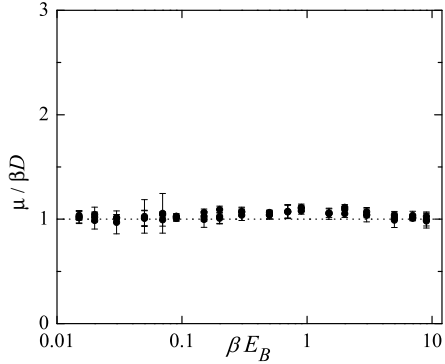}
\caption{Plot of $\mu/(\beta D)$ against $\beta E_B$ for the nine systems used in the previous figures. Six temperatures are represented for each system. \label{fig:EinsteinRelation}}
\end{center}
\end{figure}

For fixed time $t$, the error term is order $F$, but in view of the metastability of the current alluded to in the introduction and further explored in the next section, one cannot expect this term to remain small uniformly in $t$. In other words, the limit as $t$ goes to infinity cannot be taken in (\ref{eq:mobfinitet}). In fact, the system does not sustain a stationary state. On the other hand, the results of \cite{spd} (see also Fig.~\ref{fig:diffusion}) strongly indicate that
\begin{equation}\label{eq:diffcst}
D:=\lim_{t\to+\infty} D(t)
\end{equation}
does exist, for all temperatures. In fact, it is clear from the numerics that the diffusive regime is reached on a short time scale. At such times, it suffices to take $F$ small enough to make the error term in (\ref{eq:mobfinitet}) small, leading finally to the following finite time version of the Einstein relation,
$$
 \mu_F(t)\sim\beta D(t)\sim \beta D,
$$
linking the numerically computed diffusion constants and mobilities. Note that, since the right hand side depends neither on $t$, nor on $F$, this means that, for such times and fields, the same must be true for $\mu_F(t)$. We tested this relation numerically for nine systems, each at six different temperatures, as shown in Fig.~\ref{fig:EinsteinRelation}.

\section{Runaway}\label{s:runaway}
\begin{figure}[t]
\center
\includegraphics[height=6cm]{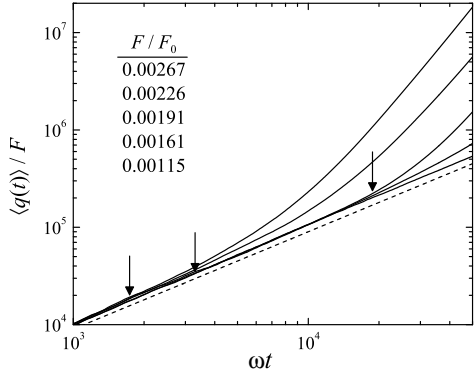}
\caption{ Log-log plot of $\langle q(t, F)\rangle/F$ against $t$ for several values of $F/F_0$, as indicated and for the parameter values $\beta E_B=0.5, E_B/E_0=5, 2\sigma/L=0.5$. The parts of the graphs parallel to the dashed line correspond to behaviour linear in time (slope equal to $1$). The common $y$-intercept with the vertical axis reflects their common value for the mobility. But for all forces, at large enough times, the displacements increase faster than linearly with time.\label{fig:runaway1}}
\end{figure}

As we have already suggested, the system studied here cannot sustain a constant drift
speed $v_{F}$  indefinitely. This can be observed by considering quite strong fields, as we have done in Fig. \ref{fig:runaway1}, in which one clearly sees the drift speed increasing rather suddenly at some critical time $t_{\mathrm c}(F)$ that depends on $F$. We will analyse this phenomenon below in order to obtain an estimate for $t_{\mathrm c}(F)$.

It is clear from (\ref{deltav}) that, on
average, particles with velocity
$$
v>\sqrt{4E_{B}E_{0}/Fa}
$$
gain more
field energy than they dissipate, so that their mean velocity increases eventually as $%
v\left( t\right) \sim v+Ft.$ In addition, one may expect that even somewhat slower particles may, as a result of thermal fluctuations, see their speed increase beyond $\sqrt{4E_{B}E_{0}/Fa}$ and then accelerate indefinitely. This suggests the definition 
\begin{equation}\label{eq:criticalspeed}
v_{\mathrm c}(F)=\gamma\sqrt{4E_{B}E_{0}/Fa}
\end{equation}
of a critical speed $v_{\mathrm c}(F)$ where $\gamma$ is a numerical constant of order $1$. Particles with speeds larger than $v_{\mathrm c}(F)$ will accelerate indefinitely, on average. Note that, obviously, this critical speed increases when the applied force decreases, and is otherwise independent of the temperature. Now, in a thermal distribution there are always particles with speeds larger than $v_{\mathrm c}(F)$ and their increasing velocities will eventually dominate the average velocity, which will no longer remain constant. Provided the field $F$ is low enough, however, the thermal speed $v_{\text{th}}=\beta ^{-1/2}$ is much lower than the critical speed $v_{\mathrm c}(F)$. Particles destined to undergo runaway are then rare, since they find themselves in the tail end of the distribution.  Indeed,
\begin{equation}\label{eq:smallforce}
v_{\text{th}}\ll v_{\mathrm c}(F)\Leftrightarrow Fa\ll \beta\gamma^2 E_{B}E_{0}
\end{equation}
and hence the fraction of particles in the thermal distribution that have a velocity $v>v_{\mathrm c}(F)$, which is given by
$$
\rho_{F} \equiv  \left({\frac{\beta}{2\pi}}\right)^{1/2}\int_{v_{\mathrm c}(F)}^{+\infty} \e^{-\frac{\beta v^2}{2}}\ \d v\sim\left( 8\pi \beta \gamma^2E_{B}E_{0}/Fa\right) ^{-1/2}\exp \left( -2\beta\gamma^2 E_{B}E_{0}/Fa\right)
$$
is exponentially small in $F^{-1}$.
Consequently,  for low enough fields most particles have speeds less than $v_{\mathrm c}(F)$ and to them the arguments of the previous sections apply, on average: they reach a limiting drift velocity $v_F$.

The average velocity of the ensemble
will, therefore, take the form%
\begin{eqnarray*}
\langle v\left( t\right) \rangle &\sim &v_{F } \left(
1-\rho_{F}\right) +\sqrt{\frac{\beta }{2\pi }}\int_{v_{\mathrm c}(F)}^{+{\infty}}\left(
p+Ft\right) e^{-\beta p^{2}/2}\ \d p \\
&\sim &v_{F} \left( 1-\rho_F\right) +\frac{1}{\sqrt{2\pi
\beta }}e^{-\frac{1}{2}v_{\mathrm c}(F)^{2}\beta }+Ft\rho_F\\
&\sim &v_F +\left( Fa\right) ^{3/2}\frac{t}{a}\frac{e^{-\beta
\gamma^2\frac{2E_{B}E_{0}}{Fa}}}{\sqrt{8\beta \gamma^2E_{B}E_{0}\pi }}
\end{eqnarray*}%

\begin{figure}[t]
\center
\includegraphics[height=6cm]{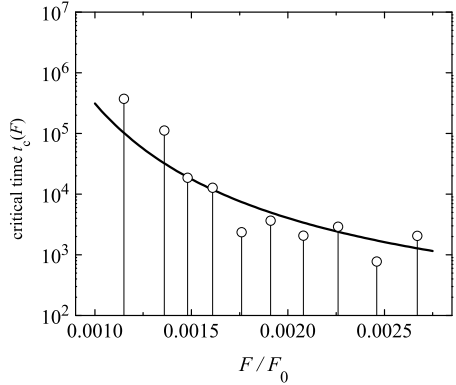}
\caption{Log-log plot of the critical breakdown times as a function of $F$ for the system with parameters $\beta E_B=0.5$, $E_B/E_0=5$, $2\sigma/L=0.5$. The circles on this graph indicate numerically established values of $t_{\mathrm c}(F)$. The full curve is
$t_{\mathrm c}(F)$ where $\gamma=0.6$ and as in the previous figures $F_{0} = a^{2} \omega$.
\label{fig:runaway2}}.
\end{figure}

Consequently, for times $t$ long enough for the slow particles to reach
their final average velocity, but short enough that the second term in this last
relation is negligible, the particles will exhibit a finite and constant current, and thus
an Ohmic response.
Thus, defining a critical time $t_{\mathrm c}(F)$ when the fast particles start to dominate
the mean particle current [i.e., when $\rho _{F}Ft_{\mathrm c}(F)\sim $ $v_{F}$], we
find that%
\begin{equation}\label{eq:runawaytime}
t_{\mathrm c}(F)=v_{F}a\sqrt{\frac{8\beta \gamma^2E_{B}E_{0}\pi }{F^{3}a^{3}}}e^{\frac{2\beta\gamma^2
E_{B}E_{0}}{Fa}}
\end{equation}%
which is exponentially long in $F^{-1}.$

We note that such an effect will be unobservable if
the length of the sample supporting the current is much shorter than the
exponentially large distance $L_{F}=v_{F}t_{\mathrm c}(F)$ the particles would drift
during this time in an infinite sample. We have tested this runaway phenomenon numerically for the system with $\beta E_B=0.5$, $E_B/E_0=5$, $2\sigma/L=0.5$.
In Fig.~\ref{fig:runaway1} one observes for this system the numerically computed value of $\langle q(t)\rangle/F$  for five values of the force $F/F_{0}$ between $1\times 10^{-3}$ and $2.5\times 10^{-3}$.   One observes that the displacement grows faster than linearly beyond a breakdown time, indicated with vertical arrows on the figure. This breakdown time is of the order $10^3 \omega^{-1} - 10^4 \omega^{-1}$ for the three largest forces used and is seen to increase with decreasing force. The numerically determined values of this breakdown time are plotted against the force $F/F_{0}$ in Fig.~\ref{fig:runaway2}. One notes that the breakdown time increases very fast with decreasing fields, in a manner qualitatively compatible with the  exponential law derived above. Indeed, the full line in Fig.~\ref{fig:runaway2} corresponds to $t_{\mathrm c}(F)$ in (\ref{eq:runawaytime}) with $\gamma=0.6$.
This is compatible also with the following observation: for much lower forces, of the order of $10^{-4} F_{0} - 10^{-5} F_{0}$,  the mean displacement of the particle increases linearly in time for the same system as seen in the circular data points of panel (c) in Fig.~\ref{fig:drifts} for times up to $t=10^7 \omega^{-1}$. This strongly indicates that for those very small forces, the breakdown time will be considerably larger than $10^7 \omega^{-1}$.

The previous considerations make clear what the difficulties and practical limits are of the numerical computations on this model, specifically at high temperatures. Indeed, because of the existence of the breakdown time, we have to use very small external forces to study the mobility (see (\ref{eq:smallforce})). But then the drift will be small and so the motion is diffusion-dominated for quite long times. Consequently, to observe the drift, one has to run many trajectories for long times in order to obtain sufficiently good statistics.

The analysis of the runaway phenomenon allows for a closer study of the long time and low field limit of (\ref{eq:mobfinitet}). First, since this is a first order formula in $F$, at fixed $t$, and in view of
(\ref{eq:diffcst}), it is clear that
$$
\mu:=\lim_{t\to+\infty}\lim_{F \to 0} \mu_F(t) =\beta D
$$
exists, with the two limits taken in that order. Due to the runaway phenomenon exchanging the limits is not possible: one cannot expect the limit as $t\to+\infty$ of the left-hand side of (\ref{eq:mobfinitet}) to exist, for any fixed value of $F$, however small. In other words, this model does not sustain a stationary state. Nevertheless, our analysis also shows that the following precise mathematical statement can be expected to be true for this model:
$$
\mu_F(t)\equiv \frac{v_F(t)}{F}=\beta D +{\mathrm O}(F) +\varepsilon(\frac1t)
$$
where the error term in $F$ is uniformly small for all times in the range
$$
\mu << t << t_{\mathrm c}(F).
$$
In other words, the following limit should exist:
$$
\mu=\lim_{\substack{F\to0, t\to+\infty \\ \mu << t << t_{\mathrm c}(F)}}\mu_F(t)=\beta D.
$$

\section{Conclusion}

In conclusion, we have shown in this paper that an array of monochromatic oscillators at a positive temperature can serve as an effective heat bath for a particle driven through the lattice under the influence of an external driving field $F$, inducing for the particle normal transport properties over very long times and distances and this in spite of the fact that a truly stationary state cannot be sustained by the system. 

To end, we point out that the situation here is therefore different from the one proven to hold for the Lorentz gas with a Gaussian thermostat \cite{cels1, cels2}. In that case a stationary state is shown to exist and the limits in (\ref{eq:mobfinitet}) can be exchanged. But there, friction is provided by a Gaussian thermostat and the model is therefore not fully Hamiltonian, as pointed out before.  On the other hand, in the Lorentz gas with rotating scatterers \cite{ey, llm}, the Einstein relation is checked numerically to hold, but for rather short chains and times. Finally, in \cite{dblp} we have argued that replacing the kinetic energy $p^2/2$ of the particle by a cosine band Hamiltonian $2V(\cos p-1)$ will suppress the runaway phenomenon, so that in that case as well a stationary state will exist.

\vfill\eject
\noindent{\bf \large Acknowledgments}

\noindent This work was supported in part by INRIA SIMPAF. P.E.P. thanks the
Laboratoire Paul Painlev\'{e} and SIMPAF for their hospitality during his
visits to the Universit\'{e} des Science et Technologies de Lille. The
numerical simulations in this paper were carried out on the
Grid'5000 experimental testbed.

\end{document}